\documentclass[prl,twocolumn]{revtex4}
\usepackage{amssymb}
\usepackage{graphicx}

\begin{document}

\title{Charged particles on a 2D plane subject to anisotropic Jahn-Teller
interactions.}
\author{T. Mertelj$^{1,2}$, V.V. Kabanov$^{1}$ and D. Mihailovic$^{1,2}$}
\date{\today}

\begin{abstract}
The properties of a system of charged particles on a 2D lattice,
subject to an anisotropic Jahn-Teller-type interaction and 3D
Coulomb repulsion are investigated. In the mean-field
approximation without Coulomb interaction, the system displays a
phase transition of first order. When the long range Coulomb
interaction is included, Monte Carlo simulations show that the
system displays very diverse mesoscopic textures, ranging from
spatially disordered pairs to ordered arrays of stripes, or
charged clusters, depending only on the ratio of the two
interactions (and the particle density). Remarkably, charged
objects with even number of particles are more stable than with
odd number of particles. We suggest that the diverse functional
behaviour - including superconductivity - observed in oxides can
be thought to arise from the self-organization of this type.
\end{abstract}

\affiliation{$^{1}$Jozef Stefan Institute, Jamova 39, 1000
Ljubljana, Slovenia}

\affiliation{$^{2}$Faculty of Mathematics and Physics, Univ. of
Ljubljana, Slovenia}

\maketitle

\newpage The standard theoretical models of
strongly correlated electrons, such as the Hubbard model\cite%
{Hubbard} or the $t-J$ model\cite{tj} neglect two important
interactions, namely long-range Coulomb repulsion and lattice
distortions caused by charged particles. Moreover, these quantum
mechanical models are typically used to study $T\approx 0$
properties. As such, these models have found limited applicability
in predicting the finite-temperature functional behaviour in
systems such as cuprate superconductors and other oxides. An
important aspect of the problem which has been of great interest
recently is the existence of intrinsic mesoscale inhomogeneity in
these systems, for
which there is mounting experimental evidence from neutron scattering\cite%
{EgamiPintschovius}, XAFS\cite{BillingeBianconi}, STM\cite{Davis} and
time-resolved carrier dynamics\cite{Demsar} amongst others%
\cite{screview}. There is emerging consensus that in doped
cuprates charge carriers may phase segregate to form nano-scale
textures. These are believed to be of importance for achieving
their functional properties, and particularly superconductivity.
The idea of charge segregation in cuprates appeared soon after the
discovery of superconductivity \cite{zaanen,emery,gorkov}, but in
most cases, long-range Coulomb repulsion was not considered. More
recently it was suggested that interplay of short range lattice
attraction and long-range Coulomb repulsion
could lead to the formation short metallic or insulating strings of polarons%
\cite{fedia1,akpz}. Since an isotropic interaction cannot lead to
stripe formation we suggested instead that an anisotropic
mesoscopic Jahn-Teller interaction between electrons and $k\neq
0~$ optical phonons might lead to the formation of pairs and
stripes\cite{mk}. A slightly different approach, based on
elasticity was considered more recently for the case of manganites
by Kugel and Khomskii \cite{klim} using the methods of Eremin et
al.\cite{eremin}, and by Shenoy et al.\cite{Shenoy}. The
importance of the interplay of long-range and short range forces
within an Ising-like model was discussed by Low et al.\cite{Low}.

The fundamental question which we try and answer here is how
charged particles order in the presence of anisotropic Jahn-Teller
type interaction, particularly when their density becomes large.
We consider charged particles on a 2D square lattice subject to
\textit{only} the long-range Coulomb interaction and an
anisotropic Jahn-Teller (JT) deformation. In the mean field (MF)
approximation without Coulomb repulsion, the system displays a
first order phase transition to an ordered state below some
critical temperature. In the presence of Coulomb repulsion global
phase separation becomes unfavorable and the system shows
mesoscopic phase separation, where the size of charged regions is
determined by the competition between ordering energy and the
Coulomb energy. Using Monte-Carlo (MC) simulations we show that
the system can form many different mesoscopic textures, such as
clusters and stripes, depending only on the magnitude of the
Coulomb repulsion compared to the anisotropic lattice attraction.
Surprisingly, a feature arising from the anisotropy introduced by
the Jahn-Teller interaction is that objects with even number of
particles are found to be more stable than with odd number
particles, which could be significant for superconductivity when
tunnelling is included\cite{mkm}.

Let us consider the JT\ model Hamiltonian\cite{mk}, and take only
the mode of $B_{1g}$ symmetry:
\begin{equation}
H_{JT}=g\sum_{\mathbf{r},\mathbf{l}}\sigma _{3,\mathbf{l}}f(\mathbf{r})(b_{%
\mathbf{l+r}}^{\dagger }+b_{\mathbf{l+r}}),
\end{equation}
where the Pauli matrix $\sigma _{3,\mathbf{l}}$ describes the
electronic doublet, $g$~is a constant, and
$f(\mathbf{r})=(r_{x}^{2}-r_{y}^{2})f_{0}(r)$ where $f_{0}(r)$
describes the effective range of the interaction\cite{mk}.

The model is reduced to a lattice gas model by using the adiabatic
approximation for the phonon field\cite{Kabanov2,klim}. The
Hamiltonian in the pseudospin ($S=1$) representation is given by:
\begin{equation}
H_{JT-C}^{LG}=\sum_{\mathbf{i},\mathbf{j}}(-V_{l}(\mathbf{i}-\mathbf{j})S_{%
\mathbf{i}}^{z}S_{\mathbf{j}}^{z}+V_{c}(\mathbf{i}-\mathbf{j})Q_{\mathbf{i}%
}Q_{\mathbf{j}}),
\end{equation}
where $Q_{\mathbf{i}}=(S_{\mathbf{i}}^{z})^{2},$ $V_{c}(\mathbf{m}%
)=e^{2}/\epsilon _{0}a m$ is 3D Coulomb potential, $%
e$ is the charge of the electron, $\epsilon _{0}$ is the static dielectric
constant and $a$ is the effective lattice constant. $S^{z}=\pm 1$
corresponds to the state with $n_{1,2}=1$, $n_{2,1}=0$, and $S_{i}^{z}=0$ to
$n_{1}=n_{2}=0$. Simultaneous occupancy of both levels is excluded due to
the large on-site Coulomb repulsion. The anisotropic short range attraction
is then given by:
\begin{equation}
V_{l}(\mathbf{m})=g^{2}/\omega \sum_{\mathbf{i}}f(\mathbf{i})f(\mathbf{m}+%
\mathbf{i}).
\end{equation}

A similar interaction can also be derived by considering the interaction of
the electronic doublet with the strain of $B_{1g}$ symmetry, taking into
account St.Venant's compatibility conditions\cite{Shenoy}. Anisotropic
attraction caused by elasticity has the form\cite{Kabanov2}:
\begin{equation}
V_{l}(\mathbf{m})=-\sum_{\mathbf{k}}\exp {(i\mathbf{k\cdot m})} \frac{g^{2}}{%
2(A_{2}+A_{1}U(\mathbf{k}))}
\end{equation}
here $A_{j}$ are the components of the elastic modulus tensor, and $U(k)=%
\frac{(k_{x}^{2}-k_{y}^{2})^{2}}{k^{4}+
8(A_{1}/A_{2})k_{x}^{2}k_{y}^{2}}$. Compared to (3), where the
range of the interaction was defined by the coupling to optical
phonons, the interaction (4) decays as $1/r^{2}$ (in 2D) at large
distances. Since these attractive forces decay faster than the
Coulomb repulsion at large distances, the net potential may have a
minimum at short distances.

Our goal is to study the model (2) at constant
average density of charged particles, $n=\frac{1}{N}\sum_{\mathbf{i}}Q_{%
\mathbf{i},}$ where $N~$is the total number of sites. However, to
clarify the physical picture we first consider a system with
a fixed chemical potential by adding the term $-\mu \sum_{\mathbf{i}}Q_{%
\mathbf{i}}$ to the Hamiltonian (2).

Models such as (2), but in \emph{the absence of the long-range
forces} were previously studied on the basis of the
molecular-field approximation \cite{lajz}.  The mean-field
equations for particle
density $n$ and pseudospin magnetization $M=\frac{1}{N}\sum_{\mathbf{i}}S_{%
\mathbf{i}}^{z}$ then have the form\cite{lajz}:
\begin{eqnarray}
M &=&\frac{2\sinh {(2zV_{l}M/{k}_{B}{T})}}{\exp {(-\mu /k}_{B}{T)}+2\cosh {%
(2zV_{l}M/{k}_{B}{T})}}  \label{eq_M} \\
n &=&\frac{2\cosh {(2zV_{l}M/{k}_{B}{T})}}{\exp {(-\mu /{k}_{B}T)}+2\cosh {%
(2zV_{l}M/{k}_{B}{T})}}
\end{eqnarray}
here $z=4$ is the number of the nearest neighbours for a square
lattice in 2D and ${{k}_{B}}${\ is the Boltzman constant}. A phase
transition to an ordered state with finite $M$ may be of either
first or second order, depending on the value of $\mu $. For the
physically important case $-2zV_{l}<\mu <0,$ ordering occurs as a
result of the first order phase transition. The two solutions of
Eqs.(5,6) with $M=0$ and with $M$ $\neq 0$ correspond to two
different minima of the free energy. The temperature of the phase
transition $T_{crit}$ is determined by the condition: $F(M=0,\mu
,T)=F(M,\mu ,T)$ where $M$ is the solution of Eq. (5). When the
number of particles is fixed (Eq.6), the system is unstable with
respect to global phase separation below $T_{crit}$.
As a result, at fixed $n$ two phases coexist with $n_{0}=n(M=0,\mu,T)$ and $%
n_{M}=n(M,\mu ,T)$, resulting in a liquid-gas-like phase diagram (Fig.1).

To investigate the effects of the long range-forces, we performed
MC simulations on the system (2). The simulations were performed
on a square lattice with dimensions up to $L\times L$ sites with
$10\leq L\leq 100$ using a standard Metropolis
algorithm\cite{Metropolis53} in combination with simulated
annealing\cite{Kirkpatrick83}\cite{mekm}. At constant $n$ one MC
step included a single update for each site with nonzero $Q_{i}$,
where the trial move consisted from setting $S_{z}=0\,$at the site
with nonzero $Q_{i}\,$and $S_{z}=\pm 1\,$at a randomly selected
site with zero $Q_{i}$. A typical simulated annealing run
consisted from a sequence of MC simulations at different
temperatures. At each temperature the equilibration phase
($10^{3}-10^{6}$ MC steps) was followed by the averaging phase
with the same or greater number of MC steps. Observables
were measured after each MC step during the averaging phase only. For $%
L\gtrsim 20$ we observe virtually no dependence of the results on
the system size.

Comparing the MC\ results in absence of Coulomb repulsion shown by $%
t_{crit}~$in Fig. 1 with MF\ theory we find the usual reduction of
$t_{crit}$ due to fluctuations in 2D by a factor of $\sim 2$, .

Next, we include the Coulomb interaction $V_{c}(r)$. We use open
boundary conditions to avoid complications due to the long range
Coulomb forces and ensure overall electroneutrality by adding a
uniformly charged background electrostatic potential (jellium) to
Eq. (2). The short range potential
$v_{l}(\mathbf{i})=V_{l}(\mathbf{i})\epsilon
_{0}a/e^{2}~$\thinspace was taken to be nonzero only for $\left\vert \mathbf{%
i}\right\vert <2$ and is therefore specified only for nearest, and
next-nearest neighbours as $v_{l}(1,0)$ and $v_{l}(1,1)$ respectively.

The anisotropy of the short range potential has a profound
influence on the
particle ordering. We can see this if we fix $v_{l}(1,0)=-1$, at a density $%
n=0.2$ and vary the next-nearest neighbour potential $v_{l}(1,1)$ in the
range from $-1$ to $1.$ When $v_{l}(1,1)<0$, the attraction is
''ferrodistortive''\ in all directions, while\ for positive $v_{l}(1,1)>0$
the interaction is ''antiferrodistortive''\ along the diagonals. The
resulting clustering and ordering of clusters at $t=0.04$ is shown in Fig.
2a). As expected, a more symmetric attraction potential leads to the
formation of more symmetric clusters. On the other hand, for $v_{l}(1,1)=1,$
the ''antiferrodistortive''\ interaction along diagonals prevails, resulting
in diagonal stripes.

In the temperature region where clusters partially order the heat
capacity ($c_L = \partial\langle E \rangle _L / \partial T$ where
$E$ is the total energy) displays  the peak at $t_{co}$. The peak
displays no scaling with $L$ indicating that no long range
ordering of clusters appears. Inspection of the particle
distribution snapshots at low temperatures (Fig. 2a) reveals that
finite size domains form. Within domains the clusters are
perfectly ordered. The domain wall dynamics seems to be much
slower than our MC simulation timescale preventing domains to
grow. The effective $L$ is therefore limited by the domain size.
This explains the absence of the scaling and clear evidence for a
phase transition near $t_{co}$.

We now focus on the shape of the short range potential which promotes the
formation of stripes shown in Fig. 2a). We set $v_{l}(1,0)=-1$ and $%
v_{l}(1,1)=0$ and study the density dependence. Since the
inclusion of the Coulomb interaction completely suppresses the the
first order phase transition at $t_{crit}$, we
measure the nearest neighbor density correlation function $g_{\rho L}=%
\frac{1}{4n(1-n)L^{2}}\sum_{\left| \mathbf{m}\right| =1}\left\langle \sum_{%
\mathbf{i}}\left( Q_{\mathbf{i}+\mathbf{m}}-n\right) \left( Q_{\mathbf{i}%
}-n\right) \right\rangle _{L}$ to detect clustering.  Here
$\left\langle {}\right\rangle _{L}$
represents the MC average. We define a dimensionless temperature $%
t_{cl}=k_{B}T_{cl}\epsilon _{0}a/e^{2}$~as the characteristic
crossover temperature related to the formation of clusters~at
which $g_{\rho L}$ rises to 50\% of its low temperature value. The
dependence of $t_{cl}$ on the density $n$ is shown in the phase
diagram in Fig. 1. Without Coulomb repulsion $V_{c}(r)$, $t_{cl}$
follows $t_{crit}$, as expected. The addition of Coulomb repulsion
$V_{C}(r)$ results in a significant decrease of $t_{cl}$ and
suppression of clustering. At low densities we can estimate the
onset for cluster formation by the temperature, $t_{0}$, at
which $g_{\rho L}$ becomes positive. It is interesting to note that $t_{0}$%
\thinspace almost coincides with the $t_{crit}$ line at low $n$
(Fig. 1).

To illustrate this behaviour, in Fig. 2b) we show snapshots of the
calculated MC particle distributions at two different temperatures
for different densities. The growth and ordering of clusters with
decreasing temperature is clearly observed. At low $n$, the
particles form mostly pairs with some short stripes. With further
increasing density, quadruples gradually replace pairs, then long
stripes appear, mixed with quadruples, etc.. At the highest
density, stripes prevail forming a labyrinth-like pattern. The
density correlation function shows that the correlation length
increases with doping, but long range order is never achieved (in
contrast to the case without $V_{c}$). Note that while locally,
there is no four-fold symmetry, the overall correlation function
still retains 4-fold symmetry.

To get further insight in the cluster formation we measured the cluster-size
distribution. In Fig. 3 we show the temperature and density dependence of
the cluster-size distribution function $x_{L}(j)=\left\langle
N_{p}(j)\right\rangle _{L}/(nL^{2})$, where $N_{p}(j)$ is the total number
of particles within clusters of size $j$. At the highest temperature $%
x_{L}(j)$ is close to the distribution expected for the random
ordering. As the temperature is decreased, the number of larger
clusters starts to increase at the expense of single particles.
Remarkably, as the temperature is further reduced, clusters of
certain size start to prevail. This is clearly seen at higher
densities (Fig.3). Depending on the density, the prevailing
clusters are be pairs up to $n\approx 0.2$, quadruples for
$0.1\lesssim n\,\lesssim 0.3$ etc.. We note that for a large range
of $v_{l}(1,0),$ the system prefers clusters with an even number
of particles. Odd particle-number
clusters can also form, but have a much narrower parameter range of stability%
\cite{mekm}. The preference to certain
cluster sizes becomes clearly apparent only at temperatures lower then $%
t_{cl}$, and the transition is not abrupt but gradual with the
decreasing temperature. Similarly, with increasing density changes
in textures also indicate a series of crossovers.

The results of the MC simulation presented above allow quite a general
interpretation in terms of the kinetics of first order phase transitions\cite%
{landau}. Let us assume that a single cluster of ordered phase with radius $%
R $ appears. As was discussed in \cite{gorkov2,mk,Kabanov2}, the
energy of the cluster is determined by three terms: $\epsilon
=-F\pi R^{2}+\alpha \pi R+\gamma R^{3}$. The first term is the
energy gain due to the ordering phase transition where $F$ is the
energy difference between the two minima in the free energy
density. The second term is the surface energy parameterized by
$\alpha $, and the third term is the Coulomb energy, parameterized
by $\gamma $. If $\alpha <\pi F/3\gamma $ $\epsilon $ has a well
defined minimum at $R=R_{0}$ corresponding to the optimal size of
clusters in the system. Of course, these clusters are also
interacting among themselves via Coulomb and strain forces, which
leads to clusters ordering or freezing of cluster motion at low
temperatures as shown by the MC simulations.

We conclude that a model with only anisotropic JT strain and a long-range
Coulomb interaction gives rise to a remarkably rich phase diagram including
pairs, stripes and charge- and orbital- ordered phases, of clear relevance
to functional oxides. The energy scale of the phenomena is defined by the
parameters used in $H_{JT-C}$ (2). For example, using the measured value $%
\epsilon _{0}\simeq 40$~\cite{LB} for La$_{2}$CuO$_{4}$, we estimate $%
V_{c}(1,0)=0.1$~eV, which is also the typical energy scale of the
''pseudogap'' in the cuprates. The robust prevalence of the paired
state in a wide region of parameters (Fig. 3 c,d) is particularly
interesting from the point of view of superconductivity. In
contrast to Bose condensation of mobile intersite bipolarons
discussed by Alexandrov and Mott\cite{AM}, it has been suggested
that pair tunnelling between objects such as shown in Fig. 2 can
lead to an insulator-to-superconductor transition \cite{mkm}. A
similar situation occurs in manganites and other oxides with the
onset of a conductive state at the threshold of percolation, but
different textures are expected to arise from the different
magnitude (and anisotropy) of $V_{l}(\mathbf{n}),$ and static
dielectric constant $\epsilon _{0}$~in the different
materials\cite{percolationmanganites}.

\begin{acknowledgments}
We wish to acknowledge valuable discussions with S.Shenoy, K.Alex Muller,
L.P.Gor'kov, A.S.Alexandrov and A.R.Bishop.
\end{acknowledgments}

\begin{figure}[h]
  \begin{center}
  \includegraphics[angle=-90,width=0.47\textwidth]{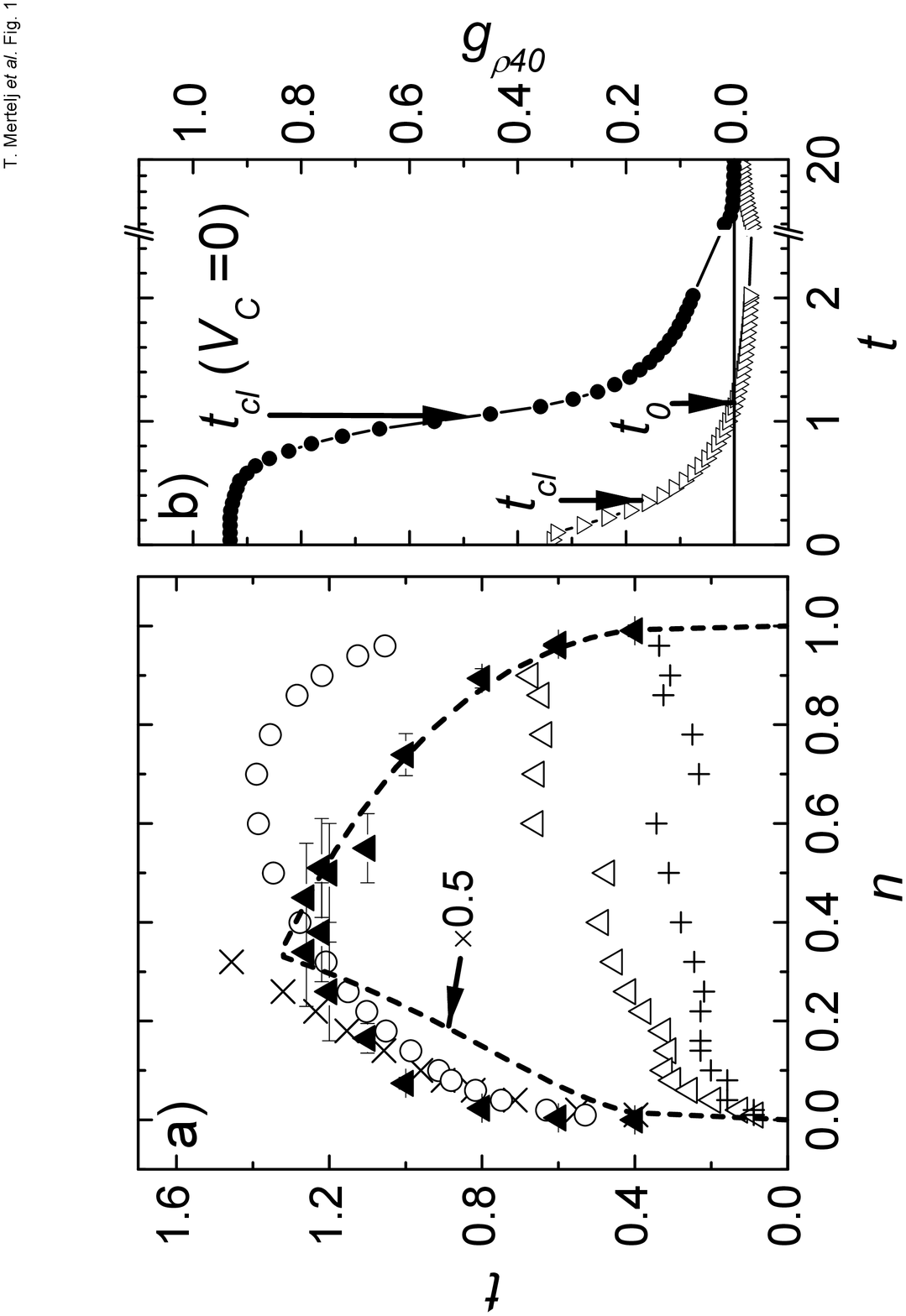}
  \end{center}
\end{figure}

Fig. 1. a) The phase diagram generated by $H_{JT}$ (2) with, and
without the Coulomb repulsion (CR). The dashed line is the MF
critical temperature,  while the full triangles ($\blacktriangle
$) represent the MC critical temperature, $t_{crit}$,
\emph{without CR}. The open circles ($\circ$) represent $t_{cl}$,
 \emph{without CR}. The open triangles ($\triangle $)\ represent
$t_{cl}$ while the diagonal crosses ($\times $) represent the
onset of clustering, $t_{0}$, \emph{in presence of }CR. The
cluster-ordering temperature (see text), $t_{co}$, (also incl. CR)
is shown as crosses (+). The size of the symbols corresponds to
the error bars. b) Typical temperature dependencies of
the nearest neighbor density correlation function $g_{\rho L}\,$for $%
n=0.18$ \emph{in absence of }CR ($\bullet $) and \emph{in presence of }CR ($%
\triangledown $). Arrows indicate the characteristic temperatures.

\begin{figure}[h]
  \begin{center}
  \includegraphics[angle=-90, width=0.47\textwidth]{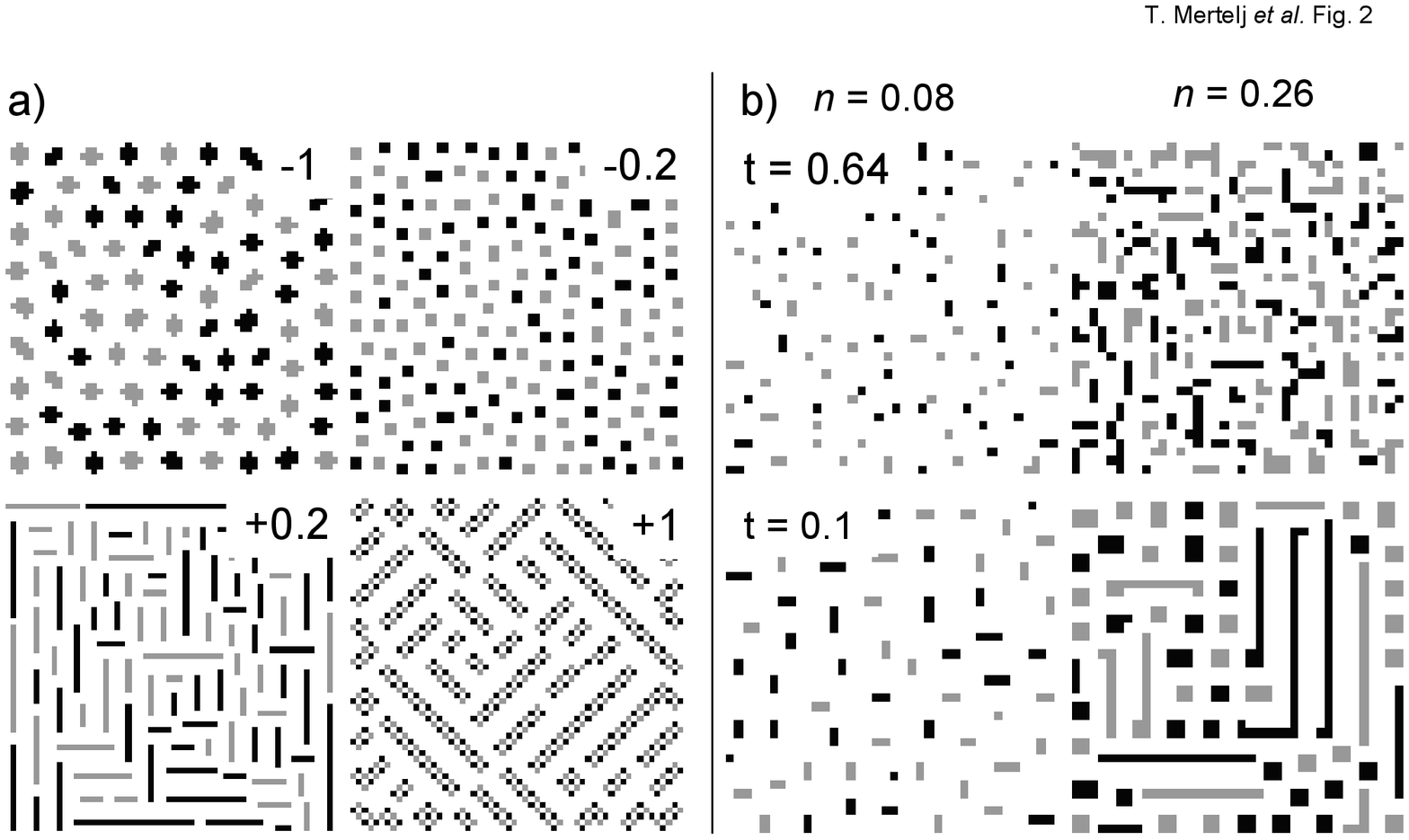}
  \end{center}
\end{figure}

Fig. 2.  a) Snapshots of clusters ordering at $t=0.04$, $n=0.2$ and $%
v_{l}(1,0)=-1$ for different diagonal $v_{l}(1,1)~$(given in each
figure). Grey and black dots represent particles clusters in state
$S_{i}^{z}=1$ and states $S_{i}^{z}=-1$ respectively. The
preference for even-particle-number clusters in certain cases is
clearly observed, for example for $v_{l}(1,1)=$ $-0.2$. b)
Snapshots of the particle distribution for two densities at two
different temperatures $t=0.64$ and $t=0.1$ respectively.

\begin{figure}[h]
  \begin{center}
  \includegraphics[angle=-90,width=0.47\textwidth]{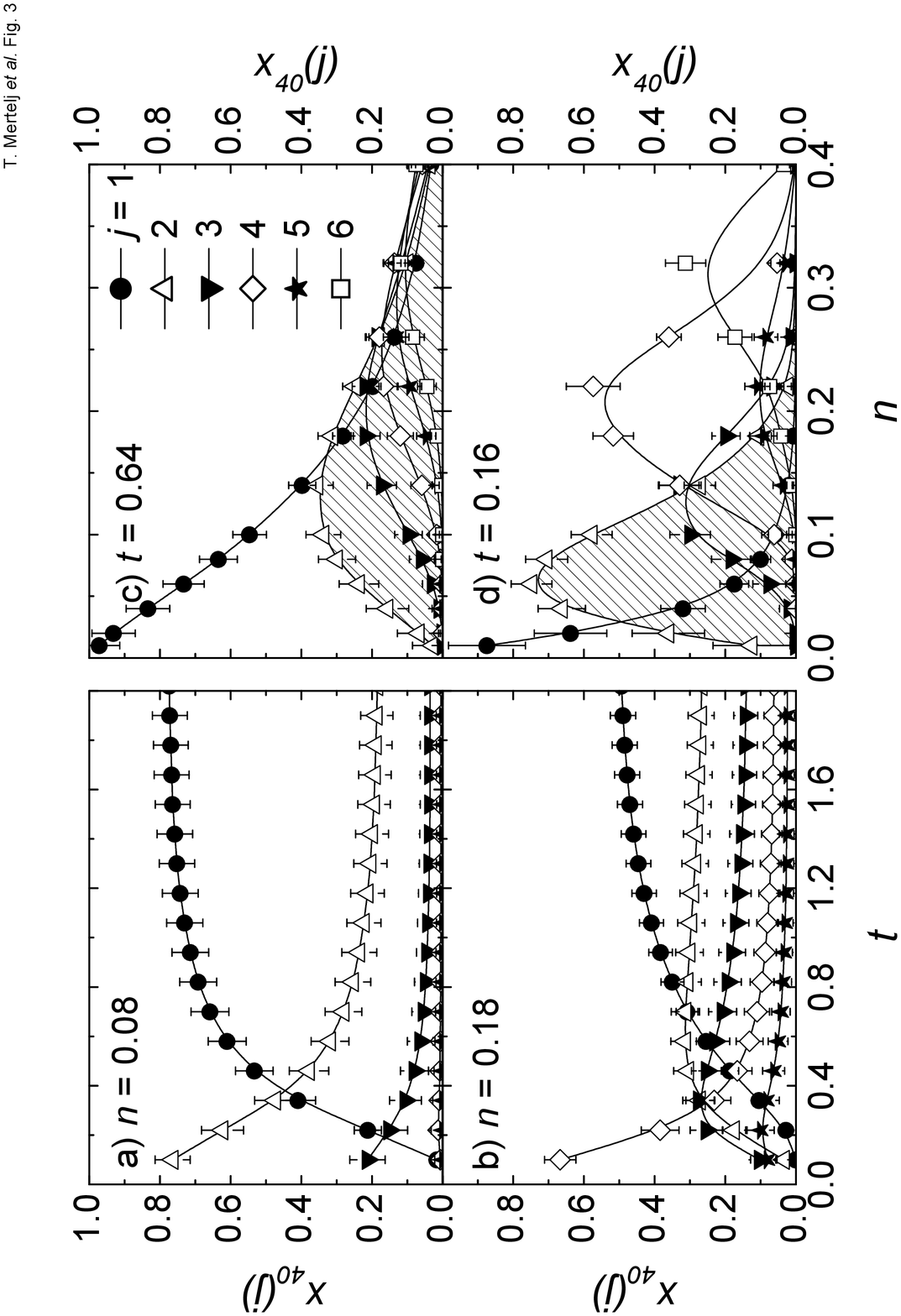}
  \end{center}
\end{figure}

Fig. 3. The temperature dependence of the cluster-size
distribution function $x_{L}(j)$ (for the smallest cluster sizes)
as a function of temperature at two different average densities
$n=0.08$ (a) and $n=0.18$ (b). $x_{L}(j)$ as a function of $n$ at
the temperature between $t_{0}$ and $t_{cl}$ (c), and near
$t_{co}$ (d). The ranges of the density where pairs prevail are
very clearly seen in (d). Error bars represent the standard
deviation.


\begin{thebibliography}{99}
\bibitem{Hubbard} E.Dagotto, Rep.Mod.Phys. \textbf{66}, 763 (1994)

\bibitem{tj} J. Jaklic and P. Prelov\v{s}ek, Adv. Phys. \textbf{49}, 1
(2000).

\bibitem{EgamiPintschovius} R.J.McQueeney et al, Phys. Rev. Lett, 82, 628
(1999).

\bibitem{BillingeBianconi} A.Bianconi et al Phys.Rev.Lett 76 3412 (1996),
Bozin et al., Phys.Rev.Lett. 84, 5856 (2000).

\bibitem{Davis} S.H.Pan et al., Nature \textbf{413}, 282 (2001), McElroy et
al., Nature \textbf{422}, 592 (2003).

\bibitem{Demsar} J.Demsar et al, Phys. Rev. Lett. 82, 4918 (1999), see also
review by D.Mihailovic and V.V.Kabanov in ``Superconductivity'',
ACS series on Structure and Bonding. Eds. A.Bussmann-Holder and
K.A.M\"{u}ller, (2004). , cond-mat/0407204

\bibitem{screview} see also: D.Mihailovic and K.A.M\"{u}ller, in High $T_{c}
$ Superconductivity 1996: Ten years after the discovery. Eds. E.Kaldis,
E.Liarokapis and K.A.M\"{u}ller, NATO ASI, Ser. E. Vol. 343 (Kluwer, 1997)
p. 243.

\bibitem{zaanen} J. Zaanen, O Gunnarsson, Phys. Rev. B, \textbf{40}, 7391
(1989).

\bibitem{emery} V.J. Emery, S. Kivelson and O.Zachar, Phys.Rev.B \textbf{56}%
, 6120 (1997).

\bibitem{gorkov} L.P. Gorkov, A.V. Sokol, Pisma ZhETF, \textbf{46}, 333
(1987).

\bibitem{fedia1} F.V. Kusmartsev, Phys. Rev. Lett., \textbf{84}, 530,
(2000), $ibid$ \textbf{84}, 5026 (2000).

\bibitem{akpz} A.S. Alexandrov, V.V. Kabanov Pisma ZhETF, \textbf{72}, 825
(2000) (JETP letters, \textbf{72}, 569 (2000)).

\bibitem{mk} D. Mihailovic, V.V. Kabanov, Phys. Rev., B, \textbf{63},
054505, (2001); V.V. Kabanov, D. Mihailovic, Phys. Rev., B,
\textbf{65}, 212508, (2002), V.V.Kabanov and D.Mihailovic,
J.Superc. \textbf{13}, 959 (2000).

\bibitem{klim} D.I. Khomskii, K.I. Kugel, Europhys. Lett. \textbf{55}, 208
(2001); Phys. Rev. B, \textbf{67}, 134401 (2003).

\bibitem{eremin} M.B. Eremin, A.Yu. Zavidonov, B.I. Kochelaev, ZhETF,
\textbf{90}, 537 (1986).

\bibitem{Shenoy} S.R. Shenoy, T. Lookman, A. Saxena, A.R. Bishop, Phys. Rev.
B, \textbf{60}, R12537 (1999); T. Lookman, et al, Phys. Rev. B, \textbf{67},
024114 (2003).

\bibitem{Low} U.L\"{o}w, V.J.Emery, K.Fabricious, S.A.Kivelson, Phys.
Rev.Lett.\textbf{72},1918 (1994).

\bibitem{mkm} D.Mihailovic, V.V.Kabanov and K.A.Muller, Europhys. Lett.
\textbf{57}, 254 (2002).

\bibitem{Kabanov2} V.V.Kabanov et al., J.Superconductivity
(subm.).

\bibitem{lajz} J. Lajzerovicz, J. Sivardiere, Phys. Rev. A\textbf{11}, 2079
(1975).

\bibitem{Metropolis53} N. Metropolis, \ et al \emph{J. Chem. Phys.} \textbf{%
21,} 1087 (1953).

\bibitem{Kirkpatrick83} S. Kirkpatrick, C.D. Gelatt and M.P. Vecchi, Science
220 (1983) 671-680.

\bibitem{landau} E.M.Lifshitz and L.P.Pitaevski, Physical Kinetics, ch.12
(Butterworth-Heinemann, 1980).

\bibitem{mekm} T.Mertelj et al. (to be published).

\bibitem{gorkov2} L.P. Gorkov, J. Supercond., \textbf{14}, 365, (2001).

\bibitem{LB} D.Reagor et al., Phys.Rev.Lett. \textbf{62}, 2048
(1989).

\bibitem{AM} A.S.Alexandrov and N.F.Mott ''Polarons and Bipolarons'', (World
Scientific, 1995).

\bibitem{percolationmanganites} E.Dagotto, T.Hotta and A.Moreo, Phys. Rep.
\textbf{344}, 1 (2001).
\end{thebibliography}
\end{document}